\documentclass[aps,pra,reprint,superscriptaddress,nofootinbib,amsmath,amssymb,amsfonts]{revtex4-2}

\usepackage[pdftex]{graphicx}% Include figure files
\usepackage{dcolumn}% Align table columns on decimal point
\usepackage{bm}% bold math% You should use BibTeX and apsrev.bst for references
\usepackage{float}
\newcommand{\lind}{\mathcal{L}}
\providecommand{\ket}[1]{|#1\rangle}
\providecommand{\bra}[1]{\langle#1|}

\usepackage{xcolor}

\newcommand{\abs}[1]{\ensuremath{\left| #1 \right|}}

\newcommand{\abst}[1]{\ensuremath{| #1 |}}

\newcommand{\norm}[1]{\ensuremath{|| #1 ||}}

\DeclareMathOperator{\argmin}{argmin}

\usepackage[USenglish]{babel}

\usepackage[bookmarks=false,pdfstartview={FitH}]{hyperref}
\usepackage[ruled,lined,algonl,boxed]{algorithm2e}

\begin{document}

\title{Compensating for non-linear distortions in controlled quantum systems}

\author{Juhi Singh\,\href{https://orcid.org/0000-0001-9807-9551}{\includegraphics[height=6pt]{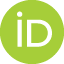}}}
\email{j.singh@fz-juelich.de}
\affiliation{Forschungszentrum Jülich GmbH,
Peter Grünberg Institute, Quantum Control (PGI-8), 52425 Jülich, Germany}
\affiliation{Institute for Theoretical Physics, University of Cologne, 50937 Köln, Germany}
\author{Robert Zeier\,\href{https://orcid.org/0000-0002-2929-612X}{\includegraphics[height=6pt]{ORCID-iD_icon-64x64.png}}}
\email{r.zeier@fz-juelich.de}
\affiliation{Forschungszentrum Jülich GmbH,
Peter Grünberg Institute, Quantum Control (PGI-8), 52425 Jülich, Germany}
\author{Tommaso Calarco}
\affiliation{Forschungszentrum Jülich GmbH,
Peter Grünberg Institute, Quantum Control (PGI-8), 52425 Jülich, Germany}
\affiliation{Institute for Theoretical Physics, University of Cologne, 50937 Köln, Germany}
\author{Felix Motzoi}
\affiliation{Forschungszentrum Jülich GmbH,
Peter Grünberg Institute, Quantum Control (PGI-8), 52425 Jülich, Germany}

%\date{\today}
%
%\makeatletter
%\def\Dated@name{}
%\makeatother
\date{May 16, 2023}

\begin{abstract}
Predictive design and optimization methods for controlled quantum systems
depend on the accuracy of the system model.
Any distortion of the input fields in an experimental platform alters the model
accuracy and eventually disturbs the predicted dynamics.
These distortions can be non-linear with a strong frequency dependence
so that the field interacting with the microscopic quantum system
has limited resemblance to the input signal.
We present an effective method for estimating these
distortions which is suitable for non-linear transfer functions
of arbitrary lengths and magnitudes provided the available training data has enough
spectral components. Using a quadratic estimation, we have successfully tested
our approach for a numerical example of a single Rydberg
atom system. The transfer function estimated from the presented method is
incorporated into an open-loop control optimization algorithm
allowing for high-fidelity operations in quantum experiments.
\end{abstract}

\maketitle
\section{\label{sec:level1}Introduction}
Over the last few decades, various quantum systems, including superconducting
circuits, neutral atoms, trapped ions, and spins \cite{Clarke:2008ul, HAFFNER2008155, Greiner:2002wt},
have shown exciting progress in controlling quantum effects for applications in
quantum
sensors \cite{Gross:2010uf}, simulators \cite{Bloch:2012ty}, and
computers \cite{Kielpinski:2002ua}.
In these setups, quantum operations are implemented using external fields or pulses
which are
generated and influenced by several electronic and optical devices.
For high-fidelity and uptime applications, this requires high performance of, e.g., population transfers
and quantum gates, while suppressing interactions with the
environment as well as decoherence. By shaping temporal and spatial profiles of external
fields and pulses, the time-dependent system Hamiltonian steers the
quantum dynamics towards the targeted outcome.

Experimental distortions of the applied pulses may reduce
the effectiveness and robustness of the desired quantum operation \cite{PhysRevA.84.022307,PhysRevLett.117.260501}.
Methods have
been developed to characterize distortions based on the
impulse response or transfer function of the experimental system
\cite{Skinner2010,PhysRevA.84.022307, PhysRevLett.110.040502,
SPINDLER201249,Kaufmann2013,Spindler2017,Rife1989419,PhysRevLett.117.260501,le2022analytic}.
These approaches for estimating field distortions
work well for distortions with a linear transfer function.
This work, however, addresses the more general case
with substantial non-linear distortions originating from the experimental
hardware.

The description of the distortions can be challenging without
knowing the exact characteristics of the experimental hardware.
Also, approximating a significant non-linearity
using a linear model will result in
model coefficients and control pulses that
are not robust against experimental distortions and suffer from a loss in fidelity. To
account for this problem, we introduce a mathematical model and an estimation method
which rely on limited experimental data
and can characterize the system behavior up to a non-linearity of finite order. To streamline our presentation, we focus
on quadratic non-linearities, but more general non-linearities
can be treated similarly.
We illustrate our estimation approach
with numerical data for a single-Rydberg atom excitation experiment in the presence of significant
non-linearities
and we highlight how our approach can calibrate for and suppress
large distortions.
We
describe an effective approach for estimating the coefficients of this non-linear
model and correct the pulses accordingly.
We emphasize that our approach is independent
of a specific experimental setup and can therefore be applied to various (spatially or temporally)
field-tunable phenomena on different quantum platforms.

Our estimation method for distortions is particularly effective in combination
with methods from quantum optimal control
\cite{Brif_2010,Glaser:2015tm,schulte2018,dalessandro2021introduction,Koch2022} and
it yields optimized pulses for highly efficient gates while accounting
for estimated distortions. To this end, we provide an analytical
expression for estimating the Jacobian of the transfer function for
quadratic distortions, which can be further generalized to higher orders.
We also validate this combined approach with our Rydberg atom excitation example.
In the context of quantum control,
any inaccuracy in the system Hamiltonian can severely affect the performance of pulses produced by optimal control.
Given a reasonably accurate model, control fields might also suffer from discretization effects,
electronic distortions, and bandwidth limitations (mostly assumed to be linear).
Accounting for these distortions by including the linear transfer
function within the dynamics, as well as its combined gradient, has been incorporated in
related optimization work
\cite{PhysRevA.84.022307,hincks2015controlling,le2022analytic,PhysRevA.88.035601,PhysRevApplied.15.034080}.
Another strategy
for minimizing non-linear pulse distortions
is to avoid high frequencies altogether in control pulses \cite{sorensen2020optimization, PhysRevA.98.022119}.

Starting from initial applications \cite{4307754,PhysRevA.37.4950},
optimal control methods
have been extensively used
in quantum computing, quantum simulation, and quantum
information processing
\cite{Glaser:2015tm,Koch2022,PhysRevB.74.161307,khani2009optimal, goerz2014robustness, doi:10.1126/science.aax9743}.
Analytic results applicable to smaller quantum systems
shape our understanding for the limits
to population transfers and quantum gates (see
\cite{Glaser:2015tm,Koch2022,Khaneja2001,KGB02,BCL+02,VHC02,Zeier2004,CHKO06,KHSY07,
YZK08,ZYK:2008,CHKO11,NZN12,CK12,VZGS14,PhysRevA.92.053414,freeman1998spin,
guery2019shortcuts,theis2018counteracting} and references therein).
Increasing the efficiency of quantum operations by numerically optimizing and
fine-tuning control parameters can rely on
open-loop or
model-based optimal control
methods \cite{Khaneja2005,Nigmatullin_2009,Tosner2009,de2011second,PhysRevA.84.022307,
Machnes2011,Bergholm2019,PhysRevA.68.062308,PhysRevLett.89.188301,goerz2019krotov}.
Our work on the estimation of distortions can be seen in the context of model-based approaches,
which might rely on an accurate gradient calculation of the analytical cost function
and thus on the knowledge of the Hamiltonian of the system \cite{Glaser:2015tm, PhysRevA.84.022307}.
This knowledge might be available in naturally occurring qubits (such as atomic, molecular, or optical systems),
but may also be estimated in engineered (solid-state) technologies.
Similarly, closed-loop (i.e.~adaptive feed-forward) control methods \cite{doi:10.1063/1.459386,GOODWIN20189,BARDEEN1997151,PhysRevA.84.022326,PhysRevLett.106.190501,
PhysRevA.92.062343,doi:10.1126/science.aax9743,M_ller_2022}
are used in situ to reduce adverse experimental effects on the control pulses,
while direct (real-time) feedback and reservoir engineering methods
can also be used where appropriate to counteract control uncertainties \cite{Solomon:2011vf,motzoi2016backaction}.

The paper is organized as follows: Section~\ref{sec:level2} sketches
the control setup for optimizing quantum experiments and describes the
conventional method for estimating the transfer function and its inclusion
in the optimization. In Sec.~\ref{sec:level3}, we detail our non-linear estimation method
using non-linear kernels. We also describe how to derive the transformation
matrix and its gradient. The non-linear effects
on quantum operations are shown with a numerical example of Rydberg atom excitations in
Sec.~\ref{sec:level4}. We apply the estimation methodology to
our numerical Rydberg example in Sec.~\ref{sec:level5} and discuss
requirements on the available measurement data.
Finally, we consider different numerical optimization methods in combination
with our estimation method
in Sec.~\ref{sec:level6} (see also Appendix~\ref{app:optim})  and conclude in
Sec.~\ref{sec:level7}. The raw data files from the simulations performed for this work are provided
in \cite{supp}.

\section{\label{sec:level2}Time-dependent Control Problems}
We aim to efficiently transferring the population from an initial quantum state to a final target state.
The evolving state of a quantum system is described by its density operator
$\rho(t)$ and the corresponding equation of motion
is written for coherent dynamics as
\begin{equation}\label{eqn:2.1}
\dot{\rho}= -i[H(t), \rho] + \lind(\rho).
 \end{equation}
The form of the Lindblad term  $\lind(\rho)$ is discussed in Sec.~\ref{sec:level4} while
the Hamiltonian can be expressed as
\begin{equation}\label{eqn:2.2}
H(t) = H_d + \sum_{i}{u_{i}(t)\, H_{i}}.
\end{equation}
The free-evolution or drift component is given by $H_{d}$, while
$H_i$ denotes the control Hamiltonians which are
multiplied with time-dependent control pulses
$u_{i}(t)$. More precisely, our goal is to transfer
a quantum system from a given initial pure state with density operator $\rho_{i}$
to a target pure-state density operator $\rho_t$ in time $T$ by varying the control
pulses $u_{i}(t)$ while minimizing the cost function
\begin{equation}\label{eqn:2.3}
C=1-\abst{\langle \rho_t | \rho(T)\rangle}^2 = 1- \abst{\mathrm{Tr}[\rho_t^ \dagger \rho(T)]}^2,
\end{equation}
where $\mathrm{Tr}(M)$ denotes the trace of a matrix $M$.
This cost function measures the difference between the target-state
density operator $\rho_t$ and the final-state density operator $\rho(T)$.
In this work, we employ gradient-based optimization methods, which are
described and discussed in Section~\ref{sec:level6}
and Appendix~\ref{app:optim}.

\begin{figure}[t]
    \includegraphics[]{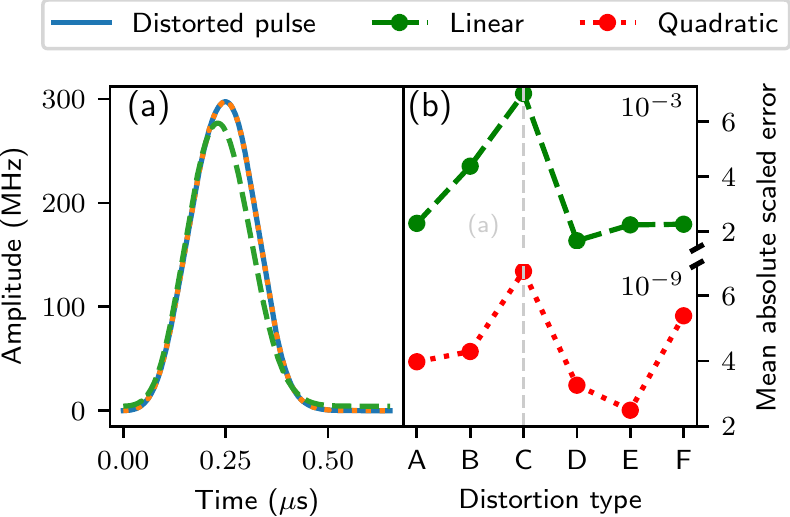}
\caption{Quadratic estimation of distorted pulses
[Eq.~\eqref{eqn:3.2}]
is preferable to linear
estimation [Eq.~\eqref{eqn:2.4}]:
(a) A pulse is numerically distorted (solid line); later the distortion is
estimated up to linear (dashed line) and quadratic terms (dotted line).
The quadratic estimation better matches the actual distorted pulse when compared to the linear estimation.
(b) Numerically computed errors for different types of distortions [including the distortion C plotted in (a)]
generated by Eqs.~\eqref{eqn:4.4}--\eqref{eqn:4.5} are plotted for both the linear and the quadratic
estimation. The error
is defined in Eq.~\eqref{eqn:5.4} and describes the difference between the actual distorted
and the estimated pulse.\label{fig:5.3}}
\end{figure}

The experimental realization of control pulses $u_{i}(t)$ relies on several devices,
which might introduce systematic distortions and reduce the overall control efficiency.
It is our objective to determine these systematic distortions
in order to adapt the control pulses during the optimization
and counteract any adverse effects.
For a linear distortion, we can calculate its transfer function
\begin{equation}\label{eqn:2.4}
T(\omega)=\frac{Y(\omega)}{X(\omega)}
\end{equation}
in the Fourier domain as the ratio of the Fourier
transform of the input and output pulses $x(t)$ and $y(t)$, i.e.\ before and
after the distortion has taken place.
Alternatively, we can calculate the impulse response $\mathcal{I}(t)$ of the system which relates
the input and output pulse in the time domain using the convolution
\begin{equation}\label{eqn:2.5}
y(t)=(x*\mathcal{I})(t) = \int_{-\infty}^{\infty}x(\tau)\,\mathcal{I}(t{-}\tau)d\tau.
\end{equation}
Figure~\ref{fig:5.3} highlights that a linear model might not be sufficient
for estimating experimental distortions as it cannot account for
non-linear effects. Non-linear
effects are demonstrated in Fig.~\ref{fig:5.3}(a) by passing one estimated example pulse
through a numerically generated
distortion [see Eqs.~\eqref{eqn:4.4}--\eqref{eqn:4.5}].
When estimating the distortion coefficients using a linear model,
the resulting distorted pulse does not match in Fig.~\ref{fig:5.3}(a)
with the actual distorted pulse. However, the quadratic estimation with
a non-linear model (as described in Sec.~\ref{sec:level3}) precisely
recovers the actual distorted pulse. Non-linear models
are, e.g., preferable for Rydberg excitations which are detailed
with realistic experimental parameters in Sec.~\ref{sec:level4}.

%%%%%%%%%%%%%%%%%%%%%%%%%%%%%%%%%%%%%%%%%%%%%%%%%%%%%
\section{\label{sec:level3}Non-linear Estimation Method }

We provide now a general approach for estimating
non-linear distortions in a controlled quantum system
and explain how this estimation approach can be incorporated
into the synthesis of robust optimal control pulses.

\subsection{Truncated Volterra series method\label{sec:level3a}}
We characterize non-linear distortions using the truncated Volterra
series method \cite{Mathews2000}. The Volterra series is a mathematical description of
non-linear behaviors for a wide range of systems \cite{Cheng2017}. In analogy to Eq.~\eqref{eqn:2.5}, we can write the general
form of the Volterra series as
\begin{equation}\label{eqn:3.1}
y(t)=h^{(0)}+\sum_{n=1}^{P}\int_{a}^{b}\!\!\!\ldots \!
\int_{a}^{b}h^{(n)}(\tau_{1},\ldots,\tau_{n})\prod_{j=1}^{n}x(t{-}\tau_{j})d\tau_{j}
\end{equation}
where $x(t)$ is assumed to be zero for $t< 0$ as we consider general, non-periodic signals.
The output function $y(t)$ can be expressed as a sum of the higher-order
functionals of the input function $x(t)$ weighted by the corresponding Volterra kernels
$h^{(n)}$. These kernels can be regarded as higher-order impulse responses of the
system. The Volterra series in Eq.~\eqref{eqn:3.1} is truncated to the
order $P<\infty$ and it is called doubly finite if
$a$ and $b$ are also finite. For a causal system, the output $y(t)$ can
only depend on the input $x(t{-}\tau_j)$ for earlier times (i.e.\ $t\geq \tau_j$)
which results in $a\geq0$; recall that $x(t{-}\tau_j)=0$ for $\tau_j>t$. The Volterra series
can therefore also model memory effects (which are assumed to be
of finite length) and it is not restricted to instantaneous effects.

The discretized form of the Volterra series truncated to second order (i.e.\ $P=2$)
is given by (\cite[Eq.~2.25]{Mathews2000})
\begin{equation}
y_n =h^{(0)}+\sum_{j=0}^{R-1}h^{(1)}_j\, x_{n{-}j}
+\sum_{k,\ell=0}^{R-1}
h^{(2)}_{k\ell}\, x_{n{-}k} \, x_{n{-}\ell},\label{eqn:3.2}
\end{equation}
The discrete output entries $y_n$ have $N$ time steps with $n \in \{0,\ldots,N{-}1\}$
which are obtained {from $L$ discrete input entries $x_q$ where $x_q=0$
for $q<0$. Note that
$N = L +R -1 \geq L$, where
$R\geq 1$} denotes the
assumed memory length of the distortion.
The memory length $R$ quantifies how the response
at the current time step depends
on the input of previous time steps, i.e., $R$ bounds the number
of previous time steps that can affect the current one.
Volterra kernel coefficients of
the zeroth, first, and second order are represented by
$h^{(0)}$, $h^{(1)}_j$, and $h^{(2)}_{k\ell}$.
The matrix given by $h^{(2)}_{k\ell}$ is symmetric.
We are characterizing the transfer function by estimating
the kernel coefficients in Eq.~\eqref{eqn:3.2}.
The number $M$ of the to-be-estimated coefficients scales quadratically with
the memory length $R$ (in general, the number of coefficients
scales with $R^P$). Although the Volterra estimation can be
extended to any higher order $P>2$, we will focus in this work on
the quadratic case.

For the estimation process, we assume that we are provided with
a training data set consisting of input-output pulse pairs $(x(t),y(t))$ from an experimental device
(or a sequence of devices) which causes the distortion. Next, we discuss how given the
training data, we can estimate the kernel coefficients in Eq.~\eqref{eqn:3.2}
by minimizing some error measures (such as the mean square error)
between the modeled output and the measured output.

\subsection{Truncated Volterra series via least squares\label{sec:level3b}}
We can choose from different methods to estimate the Volterra series.
The most widely used ones are the crosscorrelation method of
Lee and Schetzen \cite{doi:10.1080/00207176508905543} and
the  exact orthogonal method of Korenberg \cite{Korenberg:1988uv}.
We choose the latter due to its simplicity and
as it does not require an infinite-length input.
We can write Eq.~\eqref{eqn:3.2} as
\begin{align}\label{eqn:3.3}
y_n&=\sum_{m=0}^{M-1}u_{nm}\,k_m
\intertext{or equivalently as the matrix equation $Y = U K$ or}
\label{eqn:3.5}
\begin{bmatrix} y_0\\ y_1 \\[-1mm] \vdots \\ y_{N-1} \end{bmatrix}
&=
\begin{bmatrix}
u_{00} & u_{01} & \cdots & u_{0,M-1}  \\
u_{10} & u_{11} & \cdots & u_{1,M-1}  \\[-1mm]
\vdots & \vdots         &  & \vdots  \\
u_{N-1,0} & u_{N-1,1} & \cdots & u_{N-1,M-1}      \\
\end{bmatrix} \begin{bmatrix} k_0 \\ k_1  \\[-1mm] \vdots \\k_{M-1}
\end{bmatrix},
\end{align}
where $K$ is defined in Eq.~\eqref{eqn:K} below.
We follow the convention that the entries
of a given matrix (or vector) $D$ are
represented by $d_{ij}$ (or $ d_i$).
Here, $n\in\{0,\ldots,N{-}1\}$ and $m\in \{0,\ldots,M{-}1\}$ where
\begin{equation}
M= 1{+} R {+} R(R{+}1)/2
\label{eqn:3.11}
\end{equation}
denotes the number
of coefficients that need to be estimated to describe
the quadratic Volterra series.
In particular, $u_{nm}$ are obtained from the input pulses
via (recall again $x_q=0$ for $q<0$)
\[
u_{nm} =
\begin{cases}
    1& \text{for }\;  m=0,\\
    x_{n{-}m{+}1} & \text{for }\;  m\in \{1,\ldots,R\},\\
    {x_{n{-}a}\, x_{n{-}b}} & \text{for }\;  m\in \{R{+}1,\ldots,M{-}1\},
\end{cases}
\]
{where $(a,b)$ with $0\leq a \leq b \leq R{-}1$ is the $(m{-}R{-}1)$th element
in the lexicographically ordered sequence from $(0,0)$ to $(R{-}1,R{-}1)$.}
As the quadratic distortion coefficients $h^{(2)}_{k\ell}$ are symmetric, only the upper (or lower)
triangular entries need to be considered. {The column vector
\begin{equation}
K=[h^{(0)},
h^{(1)}_0,\ldots,h^{(1)}_{R-1}, h^{(2)}_{00},\ldots,h^{(2)}_{R-1,R-1}]^T
\label{eqn:K}
\end{equation}
consists of all the Volterra kernels, where $k_m = h^{(2)}_{ab}$ for
$R{+}1 \leq m \leq M{-}1$ and $(a,b)$
is chosen as above.

The example of $R=2$, $L=3$, $N=L+R-1=4$, and  $M=6$
results in (with $x_q=0$ for $q<0$)}
\begin{align}
\begin{bmatrix} y_0\\ y_1 \\y_2 \\ y_{3} \end{bmatrix}
&=
\begin{bmatrix}
1 & x_0 & x_{-1} & x_{0}x_{0}&x_{0}x_{-1}&x_{-1}x_{-1}  \\
1 & x_1 & x_{0} & x_{1}x_{1}&x_{1}x_{0}&x_{0}x_{0}\\
1 & x_2 & x_{1} & x_{2}x_{2}&x_{2}x_{1}&x_{1}x_{1} \\
1 & x_3 & x_{2} & x_{3}x_{3}&x_{3}x_{2}&x_{2}x_{2} \\
\end{bmatrix} \begin{bmatrix} h^{(0)} \\ h^{(1)}_0  \\ h^{(1)}_1  \\h^{(2)}_{00}\\ h^{(2)}_{01}\\h^{(2)}_{11}
\end{bmatrix}.
\end{align}

For the estimation of the distortions, we need to determine the values of $K$
by solving the matrix equation \eqref{eqn:3.5} with the method of least squares.
We assume now that the output data vector $Y$ has been measured in an experimental setup.
We can also concatenate multiple output pulses into a single vector to form $Y$,
which allows us to perform the estimation using multiple short pulses with different
characteristics as compared to a single long pulse. This provides the freedom of choosing
the format for our training data while observing experimental constraints.
In addition to taking a single long pulse or a set of short pulses,
we can also repeatedly use the same set of pulses to reduce the measurement error.

As the matrix $U$ contains higher-order terms of the input $x_n$, different
columns of $U$ are highly correlated with each other. This leads to the problem
of solving a linear regression model with a correlated basis set, i.e., the input variables are dependent
on each other. The precision of the estimation is adversely affected and less robust
when naively applying the method of least squares to solve the matrix equation \eqref{eqn:3.5}.
We resolve this problem by first orthogonalizing
the columns of the matrix $U$. The orthogonalization transforms the input variables (stacked in columns of $U$)
such that they are independent of each other. After orthogonalizing $U$ to $V$,
Eq.~\eqref{eqn:3.5} is transformed to
\begin{equation}\label{eqn:3.6}
Y=VW.
\end{equation}
Now we can solve the modified matrix equation \eqref{eqn:3.6}
using the method of least squares to robustly obtain the values of the vector $W$.
Finally, if the Gram-Schmidt method is
used for orthogonalization, then one can convert $W$ to $K$ by recursive methods (as explained in \cite{Korenberg:1988uv})
to extract the Volterra kernels $h^{(0)}$, $h^{(1)}_{j}$, and $h^{(2)}_{k\ell}$. In this work, we use the
QR factorization method which directly provides the values for $K$ \cite{HornJohnson:1985,GolubVanLoan:1996}.

\subsection{Gradient of the input response function\label{sec:gradientdistortion}}

Assuming that we have successfully estimated the transfer function,
we want to include this information in our gradient-based optimization.
This would allow us to also go beyond
the piecewise-constant control basis of GRAPE by
including arbitrarily deformed controls, generalizing further along the
lines of Ref.~\cite{PhysRevA.84.022307}.
We provide now an analytic expression for the corresponding gradient
(i.e.\ Jacobian) to build upon the earlier work discussed in Appendix~\ref{app:optim}.

We apply the commutativity of the convolution (i.e.\ $f*g=g*f$), e.g., by
changing the integration variable from $\tau$ to $z=t-\tau$ in Eq.~\eqref{eqn:2.5}.
Using a slight generalization, Eq.~\eqref{eqn:3.2} can be rewritten
as\footnote{Note that using Eq.~\eqref{eqn:3.7} for the estimation in
Secs.~\ref{sec:level3a}-\ref{sec:level3b} would require
a number of  coefficients given by $N\times M$ instead of only $M$ and is therefore not recommended.}

\begin{equation}
y_n = h^{(0)} + \sum_{j=0}^{L-1} h^{(1)}_{n{-}j}\, x_j
+\sum_{k,\ell=0}^{L-1}  h^{(2)}_{n{-}k,n{-}\ell}\, x_{k}\, x_{\ell},
\label{eqn:3.7}
\end{equation}
where the upper summation bound $L{-}1$ differs from $R{-}1$ in Eq.~\eqref{eqn:3.2},
i.e.~integrating over the length of the input instead of the length of the kernel.
From Eq.~\eqref{eqn:3.7}, we specify for each time step (indexed by $n$) a
scalar $K^{(0)}=h^{(0)}$, a column vector $K^{(1)}_n$ with
entries $[K^{(1)}_n]_j=h^{(1)}_{n{-}j}$, and a
matrix $K^{(2)}_n$ with entries $[K^{(2)}_n]_{k\ell}=h^{(2)}_{n{-}k,n{-}\ell}$ for
$j, k,\ell \in \{0,\ldots,L{-}1\}$. With this notation, we can write Eq.~\eqref{eqn:3.7}
as a matrix equation
\begin{equation}\label{eqn:3.8}
y_n=K^{(0)} + {X}^T K^{(1)}_n+{X}^T K^{(2)}_n {X},
\end{equation}
where the column vector $X=(x_0,\ldots,x_{L-1})^T$ has length $L$.
The corresponding partial derivatives are
given by
\begin{equation}\label{eqn:3.9}
\frac{\partial {y_n}}{\partial {X}}= K^{(1)}_n+ {X}^T [K^{(2)}_n+(K^{(2)}_n)^T],
\end{equation}
which simplifies for a symmetric quadratic kernel to
\begin{equation}\label{eqn:3.10}
\frac{\partial {y_n}}{\partial {X}}= K^{(1)}_n+ 2{X}^T K^{(2)}_n
\end{equation}
We can calculate
$\partial {y_n}/\partial{X}$ for all $n$ and then determine the Jacobian.
Eventually, the gradient of the cost function \eqref{eqn:2.3} is obtained
using the chain rule as, e.g., in \cite{PhysRevA.84.022307} and as discussed in Appendix~\ref{app:optim}.

\section{\label{sec:level4}Non-linear distortions during Rydberg excitations}

We illustrate our scenario of non-linear distortions during controlled quantum dynamics with robust state-to-state
transfers in a single Rydberg atom experiment. In recent years, Rydberg
atoms have been proven to be a promising platform for quantum
simulation \cite{Weimer:2010vq} and quantum
computation \cite{RevModPhys.82.2313}. One of the most distinctive features of
these atoms in quantum experiments is their strong and tunable
dipole-dipole interactions  \cite{Jau:2016tg,Browaeys_2016}.
For larger Rydberg atom arrays as for quantum simulators, excitation
protocols (and more general operations)
from the ground state to the Rydberg state are crucial.
We consider a gradient-based optimization of control pulses
(without feedback) for tailored
excitation pulses as outlined in Sec.~\ref{sec:level2}, (see also
Sec.~\ref{sec:level6}
and Appendix~\ref{app:optim}).

\begin{figure}[t]
\includegraphics{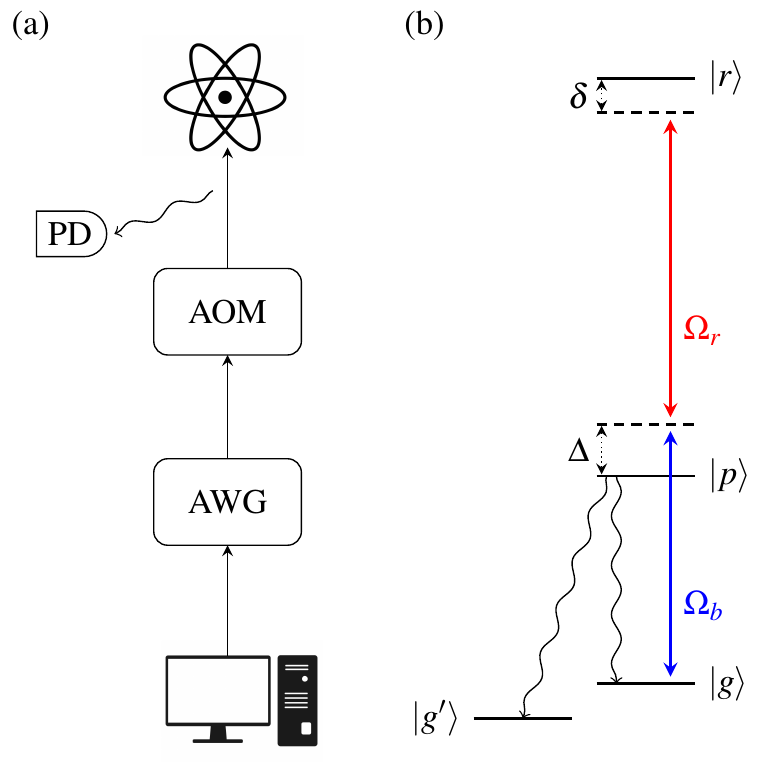}
\caption{(a) Path of the input control pulse
from the computer code via an arbitrary waveform generator (AWG)
and an acousto-optic modulator (AOM) to the atom. Before the atom,
the output control pulse can be measured using a photodiode (PD).
(b) Energy diagram for the excitation of a single Rydberg atom (see text).
\label{fig:4.2}}
\end{figure}

The Lindblad master equation for the time evolution of the system is given by Eq.~\eqref{eqn:2.1}.
Following \cite{PhysRevA.97.053803}, the model Hamiltonian for a single
Rydberg atom is equal to
\begin{align}
H(t) &= \Omega_b(t) \frac{\ket{g}\bra{p}+\ket{p}\bra{g}}{2}+
\Omega_r(t) \frac{\ket{p}\bra{r}+\ket{r}\bra{p}}{2} \nonumber \\
&- \Delta \ket{p}\bra{p} - \delta \ket{r}\bra{r}.
\label{eqn:4.2}
\end{align}
The Rabi frequency $\Omega_b(t)$ of the blue laser excites the atom from the
ground state $\ket{g}$ to the intermediate state $\ket{p}$ and the Rabi frequency $\Omega_r(t)$
excites the atom from $\ket{p}$ to the desired Rydberg state $\ket{r}$ (see Fig.~\ref{fig:4.2}(b)).
In terms of Eq.~\eqref{eqn:2.2}, $\Omega_b(t)$ and $\Omega_r(t)$ constitute
time-dependent control pulses (such as given by STIRAP \cite{stirap}).
Moreover, $\Delta$ and $\delta$ are the single-photon and the
two-photon resonance detuning, which will be for simplicity assumed to be zero
($\Delta=0$ MHz and $\delta=0$ MHz).
The Lindblad operator \cite{Lindblad:1976} reads as \cite{PhysRevA.97.053803}
\begin{equation}\label{eqn:4.3}
{\lind(\rho) =\sum\limits_{j\in\{d,g,g'\}}
(V_j \rho V_j^{\dagger}) -\tfrac{1}{2}(V_j^{\dagger}V_j  \rho + \rho V_j^{\dagger} V_j)}
\end{equation}
where $V_g = \sqrt{\Gamma_{g}} \ket{g}\bra{p}$,
$V_{g'}= \sqrt{\Gamma_{g'}} \ket{g'}\bra{p}$,
and $V_d = \sqrt{\Gamma_{d}} \ket{r}\bra{r}$ are the Kraus operators.
Here,
$\Gamma_{g}={\Gamma}/{3}$ and $\Gamma_{g'}={2\Gamma}/{3}$
denote the probability for spontaneous emission from $\ket{p}$ to the
ground state $\ket{g}$ or to $\ket{g'}$ which represents all other ground-state sublevels.
Realistic experimental parameters $\Gamma=2\pi\times 1.41$ MHz and
$\Gamma_d=2\pi\times 0.043$ MHz have been provided by the Browaeys group,
where  $\Gamma_d$ is the Doppler effect.
In a real experiment, the gradients of the controls are restricted
due to bandwidth limitations.
In particular, the controls cannot have derivatives larger than a certain
rise speed given by the experimental setup.
In our simulations, we take realistic values for the rise times of $0.1\mu s$ and $0.15 \mu s$ for
the red and blue laser pulses respectively (which translate into rise speeds).

Let us now discuss how systematic distortions can be introduced in this experimental
platform during the processing and forwarding of the control signals which finally act
on the atom(s). The path of the control signals is sketched in Fig.~\ref{fig:4.2}(a).
Starting from some computer program, the input pulse (modulated with a fixed carrier
frequency) is passed through an arbitrary waveform generator (AWG) to produce
the radio-frequency pulse. This pulse is then used as an input for an
acousto-optic modulator (AOM) which modulates the intensity of a laser beam.
The final laser pulse is then applied to the atom(s) to perform the excitation.
The AOM can shape pulses using optical effects such as
dispersion \cite{doi:10.1063/1.2409868,doi:10.1063/1.5020796}.
In this experimental setup, one can measure the laser signal before
it acts on the atom(s) using a photo diode. In summary, one can choose the input
pulse and measure the output pulse; multiple measured input-output pulse
pairs serve as training data, which is used to determine systematic distortions.

\begin{figure*}
\includegraphics{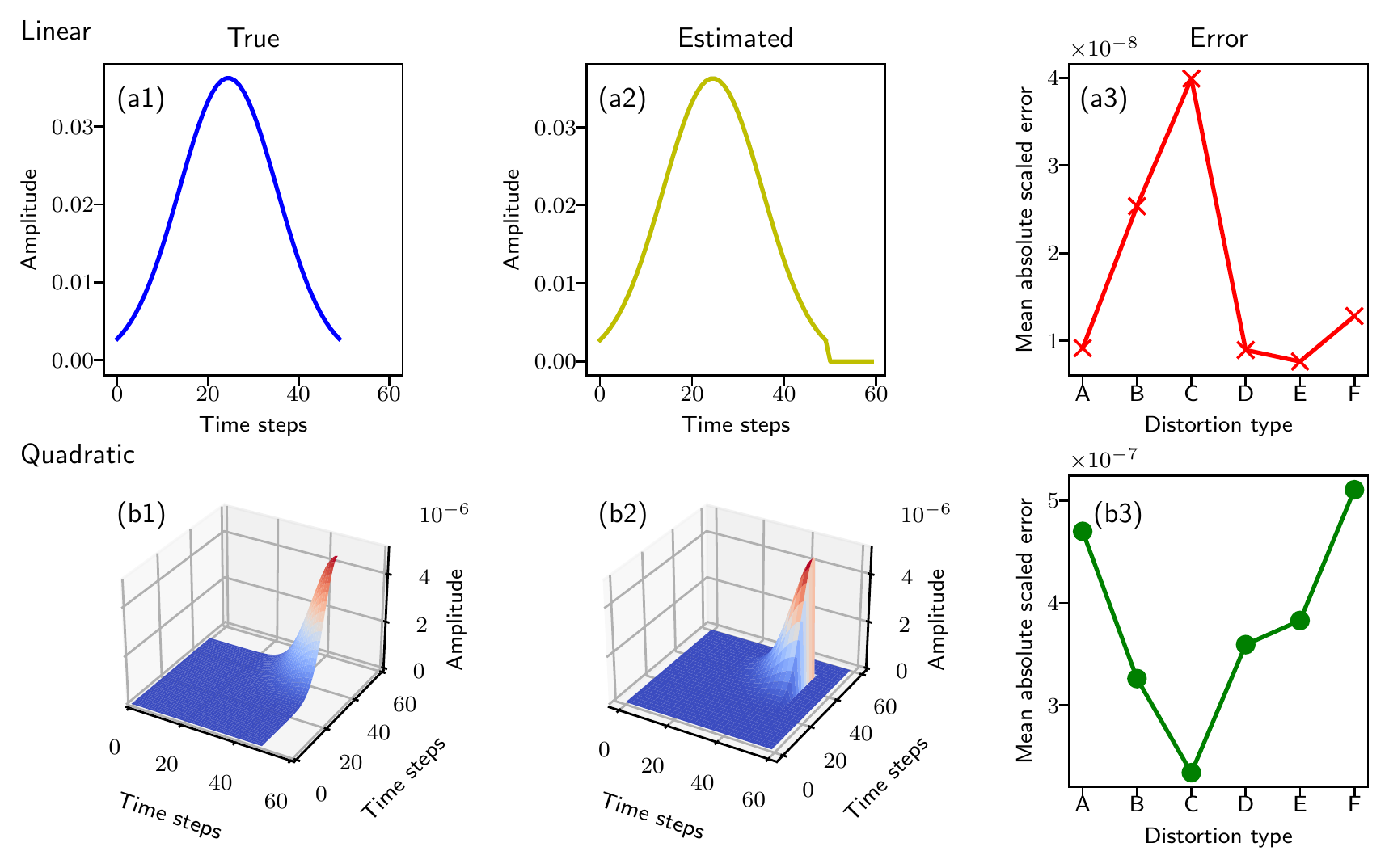}
\caption{Estimation of both the linear and quadratic components for a non-linear distortion: (a) The linear component (a1) of the distortion C
is compared with its estimated value (a2). The amplitude and
time steps are dimensionless. (a3) The mean absolute scaled error
[as defined in Eq.~\eqref{eqn:5.4}]
between the actual and the estimated values
is calculated for various types of distortions A-F [see Eqs.~\eqref{eqn:4.4}--\eqref{eqn:4.5}].
(b) Quadratic component similar as in (a).
\label{fig:5.2}}
\end{figure*}

In our simulation, we excite the Rydberg atom using the system
Hamiltonian from Eq.~\eqref{eqn:4.2} by applying our optimized
input control pulses. After
that, we introduce quadratic distortions to the control pulses and
repeat the simulation. The discrete linear and quadratic distortions are
prepared from Gaussian distributions described by
\begin{align}\label{eqn:4.4}
h_{1}(t) &= \frac{1}{\sigma\sqrt{2\pi}}\exp[{-\tfrac{(t{-}\mu)^2}{2{\sigma_{1}}^2}}], \\
\label{eqn:4.5}
h_{2}(t_{1},t_{2}) & =
J\exp[{-\frac{(t_{1}{-}\mu_{1})^2+(t_{2}{-}\mu_{2})^2}{2{\sigma_{2}}^2}}].
\end{align}
The memory length of the discretized dimensionless distortion is $R$.
For the distortions A, B, and C, we have chosen $R=50$,
standard deviations $\sigma_1$
of $1$, $6$, and $11$, and $\sigma_2$ of $4.25$, $6.37$, and $8.50$.
Similarly, for the distortions D, E, and F, we have varied $R$ between
$20$, $40$, and $60$ while fixing
$\sigma_{1}=1$ and $\sigma_{2}=4.25$. The amplitude term $J$ has been
kept  constant at $5 \times 10^{-6}$ in all cases.
The example distortion C is shown in Figs.~\ref{fig:5.2}(a1) and \ref{fig:5.2}(b1).
Throughout this work, the zeroth order kernel is set to $h^{0}=0.1$.

\begin{figure}
\includegraphics{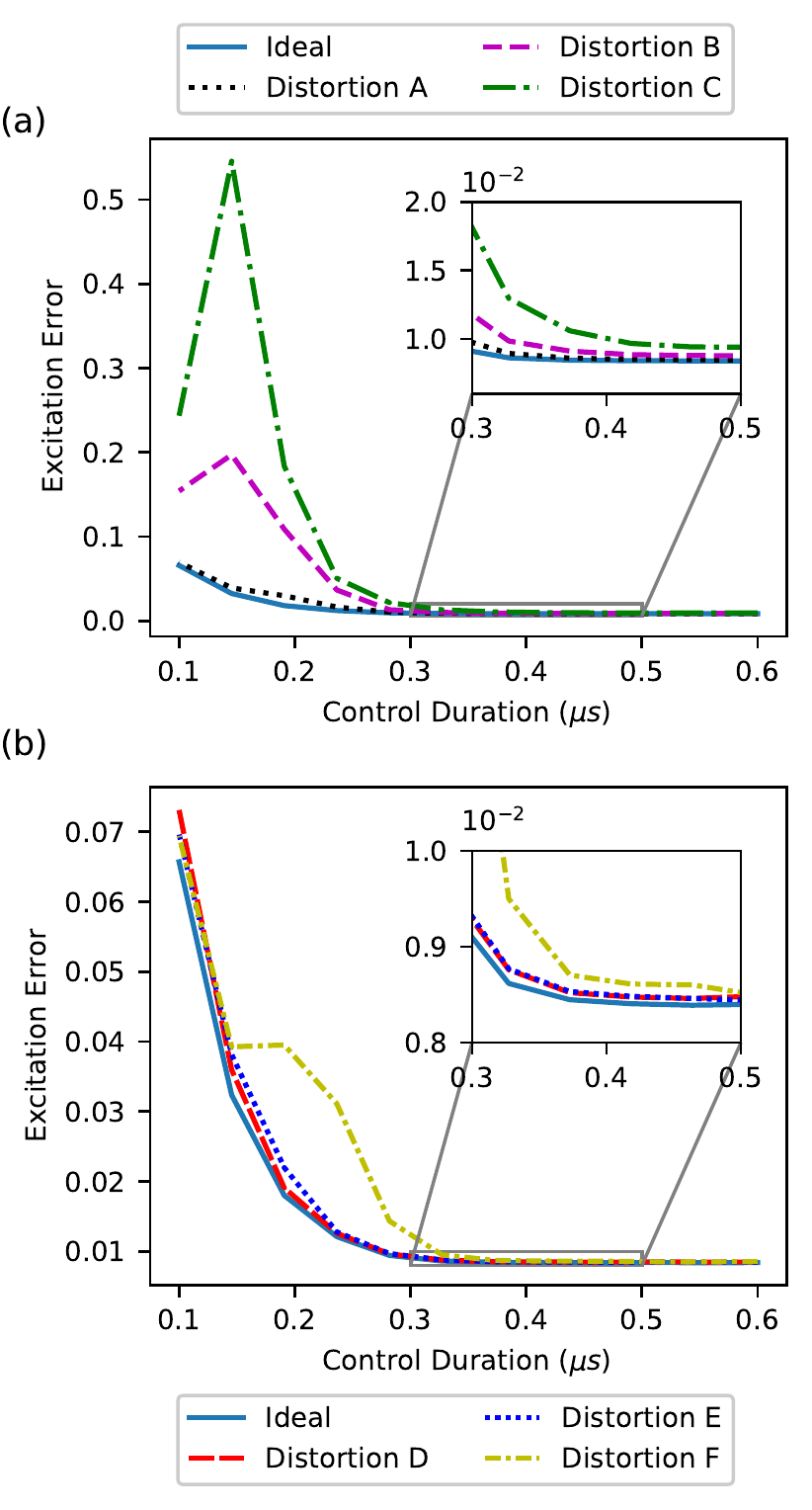}
\caption{Reduced excitation efficiencies of optimized
control pulses
due to non-linear distortions in a simulated
single Rydberg atom for distortions with
(a) an increasing variance but constant memory length (A-C) and
(b) a constant
variance but increasing memory length (D-F); refer to Sec.~\ref{sec:level4}.\label{fig:4.1}}
\end{figure}

We observe optimized controls with a simulated
Rydberg excitation error in the range from $0.06$  to $0.008$ for
different pulse lengths (see Fig.~\ref{fig:4.1}). As expected, longer total durations
for the excitation lead to smaller simulated errors. But longer pulse durations
might lead to further decoherence effects in the experimental implementation
(particularly when combined with additional experimental steps).
We, therefore, aim at reducing the length of the pulses
(e.g.\ to a pulse duration around $0.3\mu s$) with reduced excitation errors.
In Fig.~\ref{fig:4.1}(a), we notice a uniform increase in the error
magnitude when we increase the standard deviation of the Gaussian kernels of
Eqs.~\eqref{eqn:4.4}-\eqref{eqn:4.5} for the distortions A to C.
The distortions result in larger excitation errors for shorter control durations.
The case of  $0.1\mu s$ is however an exception, where the distorted pulse incidentally has a lower excitation error
when compared with the duration of $0.15 \mu s$.
The standard deviation is kept constant in Fig.~\ref{fig:4.1}(b), but we increase
the memory length for the distortions D to F which also results
in a larger excitation error. 
Similar to Fig.~\ref{fig:4.1}(a), shorter pulses result in higher excitation errors in Fig.~\ref{fig:4.1}(b),
where the excitation error for the distortion F for the durations $0.15\mu s$ and $0.2\mu s$ are coincidentally equal.
The increased excitation errors suggest that optimized control pulses
would be susceptible to distortions when applied in the Rydberg atom
experiments (and particularly for short pulse lengths). In Sec.~\ref{sec:level5},
we present estimation results building on Sec.~\ref{sec:level3}
for the considered types of distortions.
%%%%%%%%%%%%%%%%%%%%%%%%%%%%%%%%%%%%%

%

\section{\label{sec:level5}Numerical estimation results}

We report in this section on different simulated estimation results
which describe the characteristics and precision of applying
the Truncated Volterra series method while also comparing
multiple types of input control pulses used in the estimation.
We also perform the optimization for a single Rydberg
excitation again by including the distortions in the algorithm.
In each analysis, the estimated results are compared with the actual ones
using the mean absolute scaled error (MASE) measure
\begin{equation}\label{eqn:5.4}
\text{MASE}=\frac{1}{N}\sum_{i=1}^{N}
\abs{\frac{z^{\text{true}}_i}{\norm{z^{\text{true}}}}-\frac{z^{\text{est}}_i}{\norm{z^{\text{est}}}}}
\end{equation}
where $z^{\text{true}}$ is the actual value, $z^{\text{est}}$ is the estimated value, and $\norm{z}$ is the
Frobenius norm of the observable $z$ of length $N$. The
MASE is numerically more stable compared to the mean relative error, which can be very large
when the measured and the actual values are very small.

\subsection{\label{subsec:level1}Estimation of distortions}

We start with the results presented in Fig.~\ref{fig:5.2}
where numerical distortions are estimated by relying on
a single randomly generated control pulse with $4000$ time steps. We apply
different distortions to the pulse and employ the resulting
input-output pulse pairs in the estimation. In order to provide
a more realistic analysis, we add an additional noise term to
the output pulse
\begin{equation}\label{eqn:5.3}
y_{\text{noise}}=y_{\text{output}}+ \tfrac{1}{\sigma\sqrt{2\pi}}
\exp[{-\tfrac{(t{-}\mu)^2}{2{\sigma_{}}^2}}],
\end{equation}
where the noise is drawn from a normal distribution
with mean $\mu=0$ and standard deviation $\sigma=10^{-4}$.
Figures~\ref{fig:5.2}(a1)-(b1) display the linear and quadratic
contribution of the distortion C. The corresponding
estimated contributions are shown in Figs.~\ref{fig:5.2}(a2)-(b2)
which match closely with values in Figs.~\ref{fig:5.2}(a1)-(b1).
The results also emphasize that provided we can measure the output pulse
accurately, we can calculate the memory length of the distortion (which is here $R=50$)
and redundant coefficients are automatically set to
zero during the estimation for
a sufficiently large $R$ (here set to $60$).
The estimation process has been repeated for
multiple distortions of type A to F and we observe in Figs.~\ref{fig:5.2}(a3)-(b3)
low estimation errors of approximately $10^{-7}$ to $10^{-8}$.
Slight variations in the estimation error for different distortions 
could be attributed to the strength of the particular distortion or numerical noise.

We now also compare the estimation method of Sec.~\ref{sec:level3}
with a linear estimation method in the time domain which relies on
a linear impulse response [cf.\ Eq.~\eqref{eqn:2.5}]. We omit here
the very similar linear estimation in the frequency domain. We again
use the distortion types A to F from Sec.~\ref{sec:level4} for this
comparison and apply them again to a random-noise pulse of $4000$ steps to obtain
input-output pulse pairs for the estimation.
Figure~\ref{fig:5.3}(a) shows the effect of the
true and estimated distortion C when applied to
an example pulse of $0.4$$\mu s$ duration. The example pulse is
stretched under the distortion to a final duration of $0.65$$\mu s$.
The linear estimation is considerably less
precise when compared to the quadratic estimation. This effect
is confirmed in Figure~\ref{fig:5.3}(b) which plots the estimation errors
for the different distortion types A-F. Naturally, this also validates that
the chosen distortion types contain some non-linearity which is not
accounted for by a linear estimation.

\subsection{\label{subsec:level2}Orthogonalization}

One important step of the estimation method is orthogonalization
and we have discussed its significance in Sec.~\ref{sec:level3}.
To further highlight the benefits of orthogonalizing the basis functionals,
we test the estimation by directly
solving the matrix equation
\begin{equation}\label{eqn:5.1}
{U^TY=U^TUK},
\end{equation}
where $U$ is the matrix of the non-orthogonalized and correlated
basis functionals, $K$ is the to-be-estimated
vector of linear
or nonlinear kernel coefficients and $Y$ is the
measured output vector.
We compare the results with coefficients we get from solving
the matrix equation \eqref{eqn:3.6} with the orthogonalized basis set.

\begin{figure}
\includegraphics{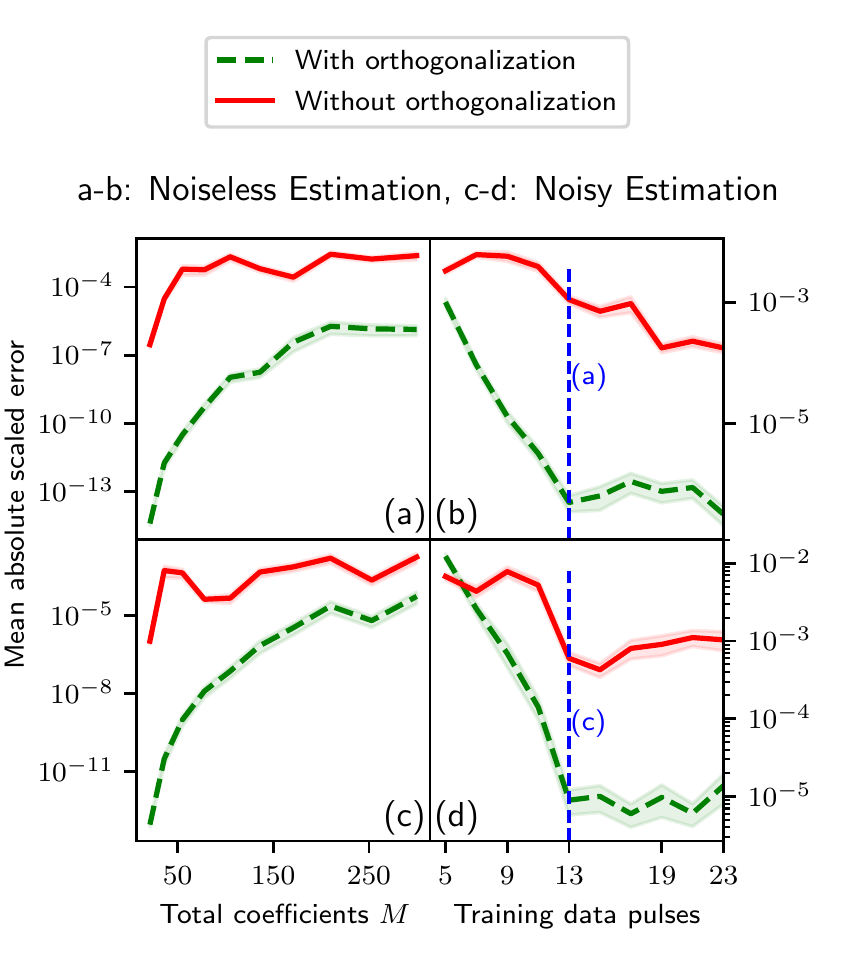}
\caption{Comparison of the simulated estimation of
non-linear distortions without and with orthogonalization
solving respectively Eq.~\eqref{eqn:5.1} and  \eqref{eqn:3.6}:
(a) Using a fixed number of noiseless training data for spline
input pulses, the relative error rises with an increasing number of coefficients
$M$. The plotted
line shows the mean error and the shaded area
indicates the spread between the $95\%$ confidence interval found from applying
the estimation results to $50$ different test pulses.
Orthogonalization is advantageous for
a larger number of coefficients. (b) Average estimation errors for different
training data sets (see text) highlight the importance of increasing the frequency
content of the available data. The averaging is performed over the full
range of all number of coefficients $M$ in (a).
(c)-(d) As in (a)-(b), but the added noise in the output pulses of the data
requires a higher frequency content for comparable error rates.
\label{fig:5.4}}
\end{figure}

In this analysis, along with the benefit of orthogonalization, we also demonstrate how the estimation depends on
the number $M$ of the to-be-estimated coefficients for the distortion,
the amount of training data, and the presence of noise in the output pulse.
Figures~\ref{fig:5.4}(a)-(b) discuss the case without added noise.
The nonlinear distortion with $\sigma_{1}=0.1, \sigma_{2}= 0.42$ and $R=5$ is estimated
using spline input pulses as the training data.
Each test and training pulse has 500 time steps and a unique frequency.
For a fixed number of spline pulses, we observe
in Fig.~\ref{fig:5.4}(a) an increasing estimation error for an
increasing number of coefficients $M$ (or memory length $R$
as $M\propto{R^2}$).
For each $M$, we apply the estimation results on
$50$ different spline pulses which serve as test data.
The corresponding mean error is plotted as a line and
the $95\%$ confidence interval is shown
as a shaded region around the mean.
In Fig.~\ref{fig:5.4}(b), we gradually increase the number of training
pulses used in the estimation.
For each fixed number of pulses, we perform the estimation on all the values of $M$
as shown in Fig.~\ref{fig:5.4}(a).
Hence each point in Fig.~\ref{fig:5.4}(b) is averaged over $500$ results.
In all cases, the estimation
benefits from being performed with orthogonalization. Also, extending
the amount of training data points by adding more spline pulses with different
frequencies improves the estimation precision as seen in Fig.~\ref{fig:5.4}(b).
For Figs.~\ref{fig:5.4}(c)-(d) in the presence
of a noise term in the output pulse with a standard deviation of $10^{-9}$, we
observe higher estimation errors which need to be compensated with
additional training data points. One can also reduce correlations
present in the training data by considering a random input pulse as its
autocorrelation is zero. However, even a completely random input pulse
results in correlations in $U$ from Eqs.~\eqref{eqn:3.5} and \eqref{eqn:5.1}
which contains various non-linear terms of the same input vector
\cite[p.~165]{Mathews2000}. In summary, Fig.~\ref{fig:5.4} illustrates the positive
effect of orthogonalization on the error rates in the estimation of the distortion.

\subsection{\label{subsec:level3}Frequency requirements}

\begin{figure}[t]
\includegraphics{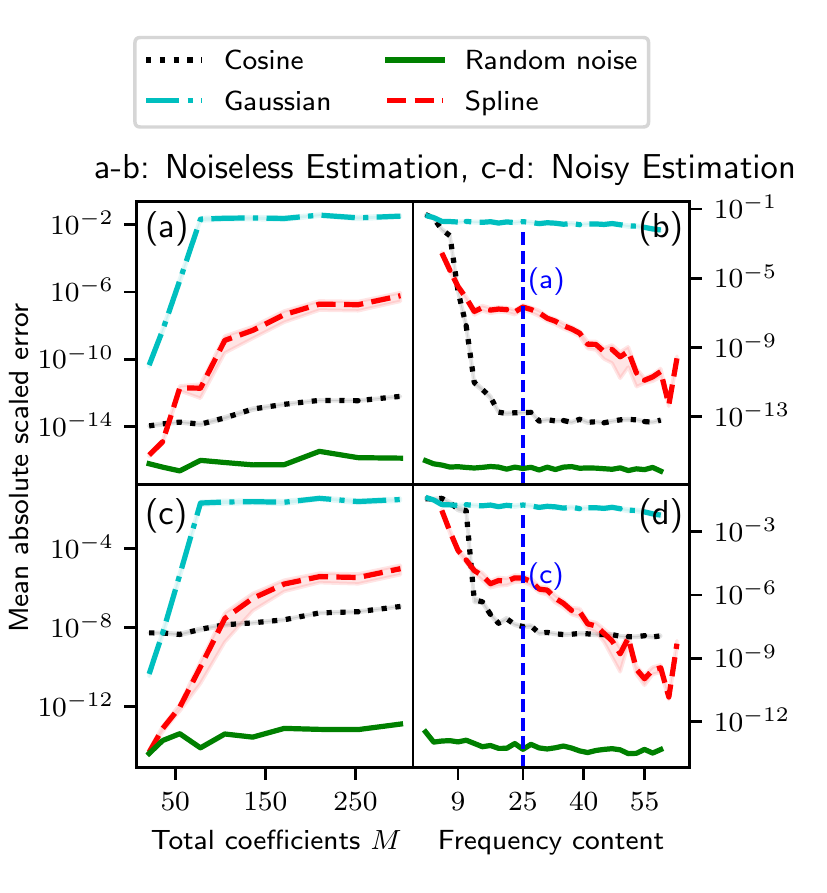}
\caption{Simulated estimation errors of distortions with multiple types
of data:
(a) training data with more frequency content (such as random-noise pulses)
perform better, even as the number of to-be-estimated coefficients $M$ increases.
(b) A lower error can be achieved  by increasing the frequency content of
the training data. For cosine and Gaussian pulses, the frequency content
is increased by adding more pulses with different frequencies, whereas
the number of knots is increased within a single pulse for splines.
Spectrally rich random-noise pulses are highly effective while keeping the data requirements low.
(c)-(d) Similar to (a) and (b), but noisy training data
increases the overall error, while random-noise pulses are the most robust.
The estimation setup is similar to Fig.~\ref{fig:5.4}.
\label{fig:5.5}}
\end{figure}

We investigate different types of training data and their performance
in the estimation following the setup of Fig.~\ref{fig:5.5}. We can order
different training data types according to their increasing frequency content,
with Gaussian pulses having the minimum frequency and random-noise pulses
having the maximum. Here, the frequency content describes the spectral content
of the training data while its value depends on the type of pulses used
(see Fig.~\ref{fig:5.4}(b) and (d).
There are different errors for spline and cosine pulses depending on the amount of data.
For a fixed number of pulses, the estimation error grows with an
increasing number of coefficients $M$ [see Fig.~\ref{fig:5.5}(a)]. Gaussian input pulses
are most strongly affected by this, while this effect is essentially negligible
in the case of random-noise pulses.
This illustrates the importance of spectrally rich input training pulses, which is further
emphasized in Fig.~\ref{fig:5.5}(b) where the estimation error is plotted, relative
to the frequency content. For different types of input pulses, the frequency
content is increased differently: we add more pulses with different standard
deviations for Gaussian pulses, we add more pulses with different frequencies
for cosine pulses, we add more
random knots to a single spline. Since a random-noise pulse has a very large bandwidth,
we aim at increasing the frequency content by increasing the number of random-noise pulses which
only slightly reduces the estimation error.
Figure~\ref{fig:5.5}(b) highlights that the frequency content is crucial for the
estimation and even a single random-noise pulse is highly effective due to its
high-frequency content.
Splines start to outperform the cosine pulses as soon as they
attain higher frequency content than the latter.
Similar conclusions hold under noise as shown
in Fig.~\ref{fig:5.5}(c)-(d) while the overall estimation error increases
for the different input pulse types when compared to the noiseless case.
The data suggests that a high-frequency content in the training pulses
might prevent overfitting noise, which is important when
working with real experimental data. Also under noise,
random-noise input pulses are most effective in the estimation due to their high frequency content.

\subsection{\label{subsec:level4}Compensating for the distortion}

\begin{figure}
\includegraphics{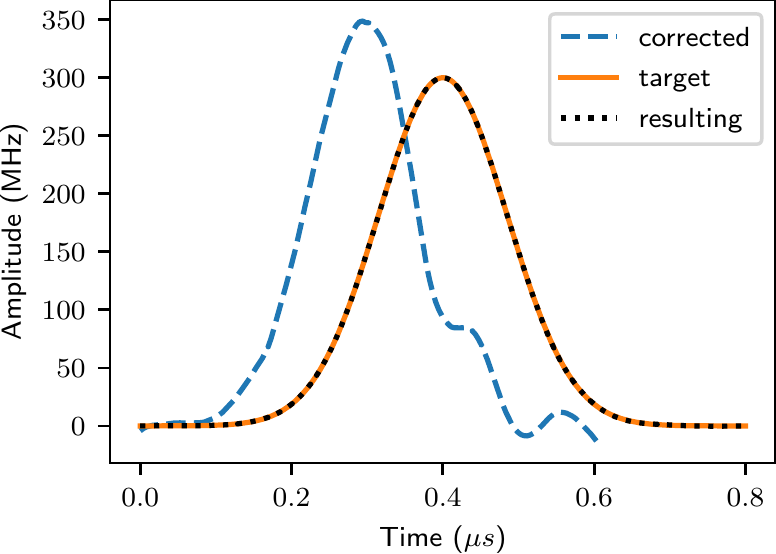}
\caption{The estimated distortion can be inverted via Eq.~\eqref{predistorted} and applied to the ideal pulse to produce a pre-distorted pulse.
The pre-distorted pulse produces the target Gaussian pulse shape after passing through 
distortion C.
\label{correct_gauss_2}}
\end{figure}

\begin{figure}
\includegraphics{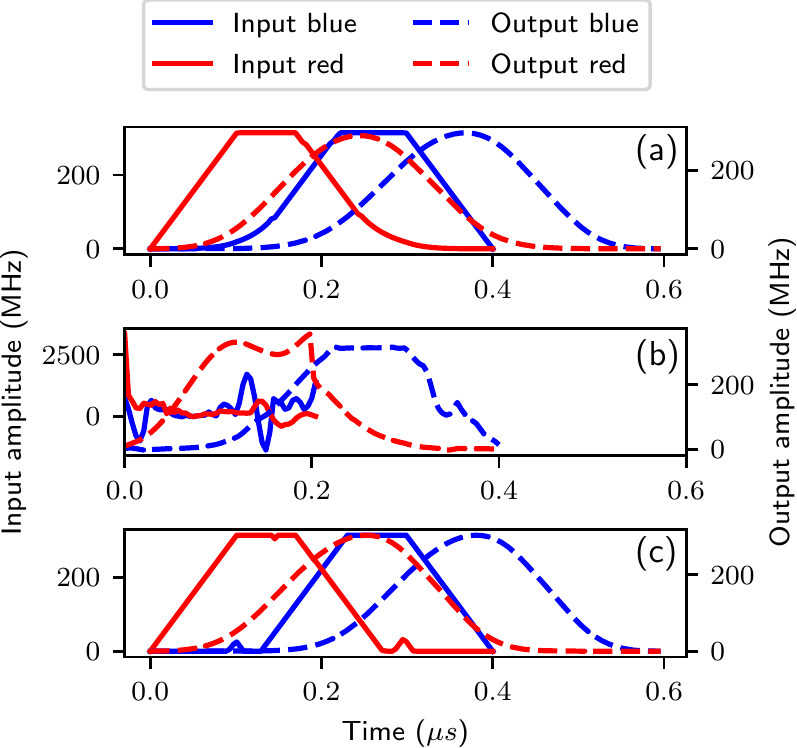}
\caption{A set of pulses before and 
after passing through the numerical distortion C in the 
control chain for the single Rydberg atom excitation. 
The blue pulse $\Omega_b(t)$ excites the atom from the
ground state $\ket{g}$ to the intermediate state $\ket{p}$ and the red pulse $\Omega_r(t)$
excites the atom from $\ket{p}$ to the desired Rydberg state $\ket{r}$ [see Fig.~\ref{fig:4.2}(b)]
(a)~Optimized input pulses and 
the corresponding outputs without correction for any distortion.
(b)~Pre-distorted pulses constructed from the distortion via Eq.~\eqref{predistorted} and the corresponding output pulses.
The output pulses do not match the optimized input pulses from (a). 
(c)~Instead of finding the pre-distorted pulses via Eq.~\eqref{predistorted}, the distortion is included in the 
optimization to reach the minimum excitation error (see Sec.~\ref{sec:level6}).
Pulses from this optimization and the corresponding outputs are shown.
\label{correct_gauss}}
\end{figure}

With the help of the estimated linear or non-linear distortion coefficients, 
we compensate for the effect of the distortion on the pulses. One natural approach to find a pre-distorted input pulse shape 
is to apply the inverted distortion to the target pulse shape. 
In the case of a linear distortion $T(\omega)$,
we would multiply $T^{-1}(\omega)$ with $X(\omega)$ [see Eq.~\eqref{eqn:2.4}]
and later transform it to the time domain. 
In the non-linear cases and for distortions expressed in time-domain kernels, 
we solve the following minimization problem to 
find the pre-distorted pulse
\begin{equation}
\argmin_{x^{\text{input}}} \frac{1}{N}\sum_{i=1}^{N}\abs{x^{\text{target}}_{i}{-}\mathcal{D}(x^{\text{input}}_{i})},
\label{predistorted}
\end{equation}
where $\mathcal{D}$ applies the distortion.
As an example, we correct one analytical Gaussian pulse in order 
to compensate for the numerical distortion C (see Figure~\ref{correct_gauss_2}). The pre-distorted
pulse constructed from the minimization problem produces the ideal Gaussian after passing through the distortion C.
Next, we perform tests on more complex optimized pulses.
As explained in Sec.~\ref{sec:level4} for the single Rydberg excitation, we have two pulses where
the blue pulse $\Omega_b(t)$ excites the atom from the
ground state $\ket{g}$ to the intermediate state $\ket{p}$ and the red pulse $\Omega_r(t)$
excites the atom from $\ket{p}$ to the desired Rydberg state $\ket{r}$ [see Fig.~\ref{fig:4.2}(b)].
Figure~\ref{correct_gauss} shows pairs of these blue and red pulses as input pulses and the
corresponding distorted output pairs. The input pulses in Figure~\ref{correct_gauss}(a)
are the ideal optimized pulses and when we do not estimate and correct
for the distortion, we receive the corresponding output pulses in the experiments.

Next, we estimate the distortion and construct the pre-distorted pulses using the minimization 
scheme discussed in Eq.~\eqref{predistorted}. The pre-distorted pulses and the corresponding
outputs after passing through the distortion are shown in Figure~\ref{correct_gauss}(b).
Unlike the analytical Gaussian pulse case (see Figure~\ref{correct_gauss_2}), we see that 
the input pre-distorted pulses do not completely reshape to the ideal optimized pulses after passing through the distortion.
This suggests that for more complex pulse shapes, pre-distorting the pulse shape with this method is insufficient to
reach a high excitation efficiency. However, we can include the estimated distortion in the optimization
to produce pulse shapes that give minimum Rydberg excitation error in the presence of the distortion.
The detailed discussion and results of this method are presented in Sec.~\ref{sec:level6}. Figure~\ref{correct_gauss}(c) shows 
an example of input pulses produced from this method and the corresponding distorted output pulses.
Note that the input-output pairs in Figure~\ref{correct_gauss}(a) and Figure~\ref{correct_gauss}(c) are quite similar.
We expect this behavior since the excitation error from these example pulses of $0.4\mu s$ duration [see Fig.~\ref{fig:5.6}]
is not much affected by the distortion.
For shorter pulses, the optimization can however produce more complex pulses different from the ideal ones in order to 
compensate for the distortion.
Therefore, we recommend to include the distortion in the optimization to compensate for the effect of the distortion
as in Fig.~\ref{correct_gauss}(c) and Sec.~\ref{sec:level6}, especially for complex pulse shapes.  

\section{\label{sec:level6}Application in optimal control}
Starting from early developments in the field, various theoretical
and experimental aspects of quantum control have been discussed
in the recent review \cite{Koch2022}.
The overall aim of quantum control is to shape a set of
external field pulses that drive a quantum system and perform a given quantum process efficiently.
While the analytical way of finding the
control parameters works for special cases, one can use
highly developed numerical tools in the context of
optimal control theory. One solves the Schrödinger or master equation iteratively and
produces pulse shapes that perform the desired time evolution.
Quantum optimal control is broadly divided into at least the two categories of
open-loop and closed-loop. Open-loop
methods can be gradient-based or not.
Open-loop control is based on the available information about the
Hamiltonian of the system and hence it suffers when the system parameters are
not completely known such as in the case of an engineered quantum system (such as solid state systems) or when
the model cannot be solved precisely as in the case of many-body dynamics \cite{PhysRevLett.109.153603}.
These limitations might be overcome by means of closed-loop optimal control where the control parameters are updated
based on the earlier measurements results \cite{PhysRevA.84.022326,PhysRevLett.106.190501}.
Closed-loop quantum optimal control can be implemented via
both gradient-based and gradient-free algorithms \cite{PhysRevLett.118.150503, PhysRevLett.112.240504, PhysRevApplied.7.041001}.
In some cases, hybrid approaches have also been suggested \cite{PhysRevLett.112.240503}.
But in the case where the system Hamiltonian is well known, open-loop control
provides more freedom to precisely tune the controls depending on experimental constraints and
generally explore a wider range of control solutions.
Moreover, it also gives a better understanding of the system and works well with systems where
fast measurements are not feasible or very noisy, in contrast to closed-loop methods which may require many
measurements to converge.

\begin{figure}
\includegraphics{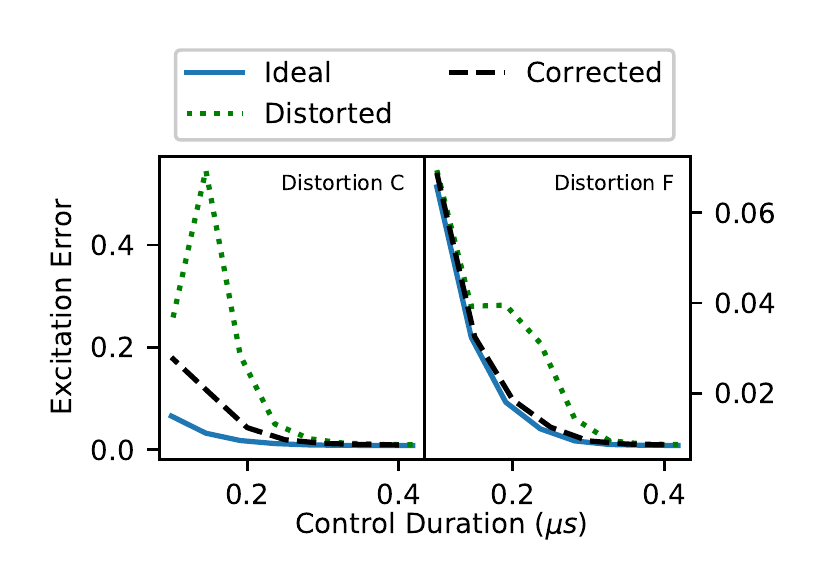}
\caption{Reduced excitation errors after correcting for the distortion with a
gradient-based optimization relying on the trust-region method.
(a) distortion C: significant reduced errors, (d) distortion F: mostly recovers
the ideal case.
\label{fig:5.6}}
\end{figure}

To take full advantage of the open-loop control method and to provide more robust pulses,
one can also characterize the experimental system completely or at least partially. Here, we highlight how
the estimation method from Sec.~\ref{sec:level3} can be employed in an open-loop control setting
to minimize the cost function $C$ in Eq.~\eqref{eqn:A.2}
by relying on the corresponding gradients as computed via
Eqs.~\eqref{eqn:3.9} and \eqref{eqn:A.9}.
We refer to Appendix~\ref{app:optim} for details.
This compensates for distortions
and decreases the error. Figure~\ref{fig:5.6} shows test minimizations of the cost function
using the trust-region constrained algorithm \cite{doi:10.1137/1.9780898719857}, which
can perform constraint minimization with linear or non-linear constraints on the control pulses.
Trust-region methods allow us to explicitly observe bandwidth limitations of the control hardware such
as limited rise speeds as discussed in Sec.~\ref{sec:level4} by enforcing the corresponding
pulse constraints.
Since distortions C and F defined by Eqs.~\eqref{eqn:4.4} and \eqref{eqn:4.5}, have the strongest effects on the Rydberg excitation error (see Fig.~\ref{fig:4.1}),
we correct the control pulses affected by them in the simulations.
We limit our test to pulses with shorter durations ranging from 0.1$\mu s$ to 0.4$\mu s$
as they are less susceptible to decoherence and hence might be more suitable for the excitation process.
We compare the excitation error
produced from the corrected pulse with the ideal and the distorted pulse excitation error.
In particular, Figure~\ref{fig:5.6} shows that the effect of the distortion C can be significantly
reduced, but it cannot be completely corrected
due to a large standard deviation and long memory length in the distortion.
The distortion F has a small standard deviation combined with a
long memory length which still produces strong effects on the control pulse but
with a generally weaker distortion. In this case, the effect of the distortion can
be almost completely corrected.

\begin{figure}
\includegraphics{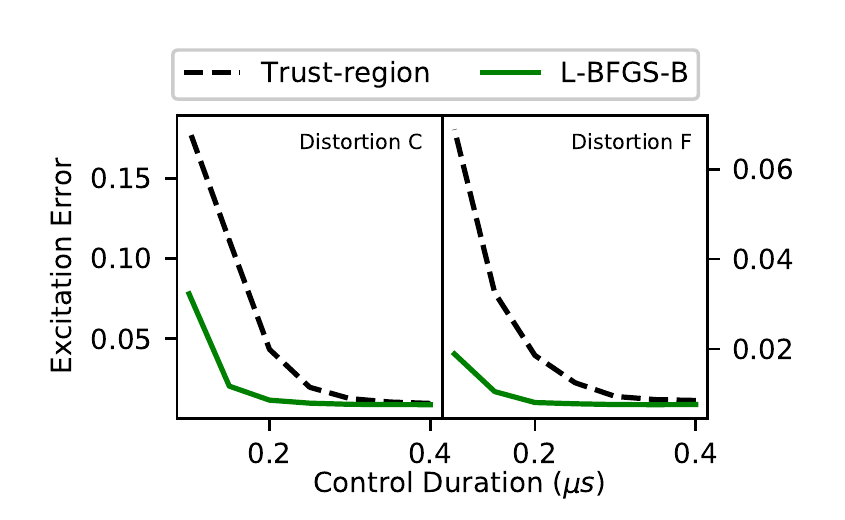}
\caption{Reduced excitation error for  L-BFGS-B when
compared to the trust-region method, even though
L-BFGS-B does not explicitly enforce constraints on
the control pulses (such as limited rise speeds).
But the correctly estimated transfer function
will implicitly account for pulse constraints.
Also, L-BFGS-B  is more effective
in the optimization.
\label{fig:5.7}}
\end{figure}

The estimation of transfer functions in order to correct for distortions has one
additional benefit. The experimental hardware given by, e.g., AWGs and AOMs usually has
bandwidth limitations which translate into limited rise speeds as discussed in
Sec.~\ref{sec:level4}. In the process of characterizing the experimental devices
via their transfer function, we also estimate the effects of these bandwidth limitations.
The estimated transfer function is then applied during the optimization,
which mirrors the effects in the experimental platform and implicitly enforces limitations
on the bandwidth or rise time.
Assuming that the bandwidth-limiting effect of the estimated transfer function
is pronounced enough,
this allows us to use the limited memory
Broyden–Fletcher–Goldfarb–Shanno (L-BFGS) algorithm to perform the
minimization of the cost function \cite{de2011second}. L-BFGS usually offers
a more efficient optimization but it cannot explicitly
account for general linear or non-linear constraints.
In the corresponding
optimizations, we only enforce simple box constraints
to limit the amplitude of the controls while using the extended L-BFGS or L-BFGS-B algorithm
\cite{53712fe04a3448cfb8598b14afab59b3}.
The results are shown in Fig.~\ref{fig:5.7} where L-BFGS-B improves the excitation efficiency
more effectively than the trust-region method (which needs to also explicitly enforce
the constraints on the rise speeds).
In summary, combining the estimation of distortions with gradient-based optimizations
can often effectively compensate for these non-linear distortions during an open-loop
optimization.

%%%%%%%%%%%%%%%%%%%%%%%%%%%%%%%%%%%%%%%%%%%%%%%%%%

\section{\label{sec:level7}Conclusion}
We have proposed a method for estimating
non-linear pulse distortions originating from
experimental hardware. Hardware limitations affect the performance
of optimal control pulses as highlighted using numerical data for
single Rydberg atom excitations. In this case, the errors are increased
for distorted control pulses beyond purely linear effects.
We provide a general model for
describing the complex characteristics of these non-linear effects.
To incorporate estimated distortions into
open-loop optimizations, we have detailed
a formula to determine
the Jacobian of the transfer function.

We tested and validated our
proposed method by
efficiently estimating different numerical quadratic distortions with varying strength
and duration. We have also shown that linear estimation methods
cannot effectively handle non-linear transfer functions. From our detailed analysis and tests,
we deduce that the orthogonalization (as described
in Sec.~\ref{subsec:level2}) is key for a robust estimation.
A robust least-squares estimation is effective only after
the orthogonalization is applied to the matrix containing the training data
as its correlated columns would otherwise interfere with the estimation.
Another critical requirement for effectively performing
the estimation is training data with enough frequency content.
Large frequency content such as in random-noise pulses better
captures the non-linear features of transfer functions, particularly in the
presence of measurement noise.

Since the estimation method is independent
of any particular type of device characteristics, it can easily be adapted to a
wide range of experimental platforms.
Combining our estimation method with
existing numerical optimization techniques can improve the quality and
robustness of quantum operations. Our work thereby addresses a key challenge
of enhancing the accuracy and robustness
of experimental quantum technology platforms.

\begin{acknowledgments}
The authors acknowledge funding from the EU H2020-FETFLAG-03-2018 under
grant agreement No 817482 (PASQuanS). We also appreciate support from the German Federal Ministry of Education
and Research through the funding program quantum technologies---from basic research to market
under the project FermiQP, 13N15891. We would like to thank
Antoine Browaeys, Daniel Barredo, Thierry Lahaye, Pascal Scholl, and Hannah Williams for
the illuminating discussions about the Rydberg system as well as providing detailed
experimental parameters. R.Z. would like to thank
Jian Cui for initial discussions about the Rydberg setup.
\end{acknowledgments}

\appendix

\section{\label{app:optim}Optimization algorithm}
We work with open-loop optimal control and detail how to incorporate
the Jacobian of the transfer function which can be determined
following Sec.~\ref{sec:gradientdistortion}.
In the mathematical statement of an optimal control problem, the fidelity function
$C$ is minimized
with regard to the control values $u_i$.
We can apply the gradient-based optimization technique known as
GRAPE \cite{Khaneja2005} which can also utilize
Newton or quasi-Newton (BFGS) methods \cite{de2011second,PhysRevA.84.022307, dalgaard2020hessian}.
We assume that the total control duration $T$ is divided into $L$ equal steps of duration $\Delta t= T/L$.
During each time step, the control amplitudes $u_i$ are constant.
The time evolution of the quantum system during the $j$th time step is given by
\begin{equation}\label{eqn:A.1}
U_j=\exp{(-i\Delta t(H_0+\sum_{i}^{}{u_i(j)H_i}))}.
\end{equation}
The cost function can be written as
\begin{equation}\label{eqn:A.2}
C=1-\abst{\bra{\rho_{t}}{{U_{L}\cdots U_{1}\rho_{i}U_{1}^{\dag}\cdots U_{L}^{\dag}}}\rangle}^2.
\end{equation}
From the inner product definition and invariance of the trace of a product under cyclic permutations of the factors,
Eq.~\eqref{eqn:A.2} can be rewritten as,
\begin{equation}\label{eqn:A.3}
1-\abst{\!\underbrace{\bra{U_{j+1}^{\dag}\cdots U_{L}^{\dag}\rho_{t}U_{L}\cdots U_{j+1}}}_{\lambda_j}
\underbrace{{U_{j}\cdots U_{1}\rho_{i}U_{1}^{\dag}\cdots U_{j}^{\dag}}\rangle}_{\rho_j}\!}^2.
\end{equation}
Here, $\rho_j$ denotes the density operator at the $j$th time step
and $\rho_{t}$ is
the backward propagated target operator at the $j$th time step.
If we perturb $u_{i}(j)$ to $u_{i}(j)+\delta u_{i}(j)$, the derivative of $C$ is given in terms of
the change in $U_j$ to the first order in $\delta u_{i}(j)$ which
is calculated by the Fréchet derivative method \cite{doi:10.1137/080716426} using the Python
package SciPy~\cite{2020SciPy-NMeth}.

In order to minimize $C$, at every iteration of the algorithm, we update the controls by
\begin{equation}\label{eqn:A.8}
u_{i}(j)\rightarrow u_{i}(j) - \epsilon\frac{\delta C}{\delta u_{i}(j)},
\end{equation}
where $\epsilon$ is a small unitless step matrix.
Next, we follow the derivation in \cite{PhysRevA.84.022307}, where
the product rule for gradient calculation is applied and one obtains
\begin{equation}\label{eqn:A.9}
\frac{\delta C}{\delta u_{i}(j)}=\sum_{n=0}^{N-1}{\frac{\delta s_k(n)}{\delta u_k{(j)}}\frac{\delta C}{\delta s_{k}(n)}}
\;\text{ where }\; \frac{\delta s_k(n)}{\delta u_k{(j)}}=T_{k}(n,j).
\end{equation}
Compared to \eqref{eqn:3.9}, $u_k$ corresponds to the input pulse $x$ and $s_n$ corresponds to the output pulse $y$.
Hence we can calculate each column of $T_{k}$ from Eq.~\eqref{eqn:3.9} as
\begin{equation}\label{eqn:A.10}
T_{k}(n)= \frac{\delta y_{n}}{\delta X},
\end{equation}
and insert $T_{k}$  into Eq.~\eqref{eqn:A.9} to calculate the effective gradient.

%\bibliography{literature}

\end{document}